\documentclass[conference]{IEEEtran}

\usepackage{cite}
\usepackage{amsmath,amsthm,amssymb}
\theoremstyle{remark}
\newtheorem*{remark}{Remark}
\usepackage{url}
\usepackage{tikz}
\usetikzlibrary{positioning, shapes.geometric, automata, calc, arrows}
\usepackage{pgfplots}
\usepackage{booktabs}

\begin{document}
%
\title{Grammatical Inference as a Satisfiability Modulo Theories Problem}

\author{\IEEEauthorblockN{Rick Smetsers}
\IEEEauthorblockA{Institute for Computing and Information Sciences\\
Radboud University\\
Nijmegen, The Netherlands\\
Email: r.smetsers@cs.ru.nl}
}


%


\maketitle

\begin{abstract}
    The problem of learning a minimal consistent model from a set of labeled sequences of symbols is addressed from a satisfiability modulo theories perspective.
    We present two encodings for deterministic finite automata and extend one of these for Moore and Mealy machines.
    Our experimental results show that these encodings improve upon the state-of-the-art, and are useful in practice for learning small models. 
\end{abstract}


%
\IEEEpeerreviewmaketitle

\section{Introduction}
One of the best studied problems in grammatical inference is that of finding a deterministic finite automaton (DFA) of minimal size that accepts a given set of positive examples and rejects a given set of negative examples of an unknown language.
This problem can be very hard.
It is the optimization variant of the problem of finding a consistent DFA of a fixed size, which has been shown to be NP-complete \cite{Gol1967}.

In \cite{CN1997}, Coste and Nicolas observe that this problem can be encoded as a graph coloring problem.
The intuition behind their encoding is as follows.
First, a tree-shaped DFA is constructed that accepts exactly the positive examples and rejects exactly the negative ones.
Each state in this DFA is represented by a vertex in a conflict graph.
Two vertices in the graph are connected by an edge if one vertex represents an accepting state and the other represents a rejecting state.
Now, the problem at hand is to color this graph, with the additional constraint that for states that are represented by vertices of the same color, their parents have to be represented by vertices of the same color as well.
For such a coloring, a minimal DFA can be constructed in which each state is represented by a different color.

In \cite{HV2013}, Heule and Verwer propose an encoding of the aforementioned graph coloring problem in propositional logic.
\emph{Satisfiability}, or SAT, is the problem of deciding if there exists an assignment to a propositional logic formula that makes it true.
To prove that the minimal size of a DFA is $n$, Heule and Verwer use an iterative procedure to determine that an encoding for $n$ colors is satisfiable, but an encoding for $n-1$ colors is unsatisfiable.

For many applications, encoding the problems into propositional logic is not the right choice.  
Frequently, a better alternative is to express the problems in a richer logic.
A first effort in this direction was made by Bruynooghe et.\ al.\ \cite{BBB2015}, who express the encoding by Heule and Verwer in a predicate logic.

In this paper we encode the aforementioned grammatical inference problem as a \emph{satisfiability modulo theories} (SMT) problem.
SMT is the problem of deciding the satisfiability of a formula with respect to one or more background theories expressed in first-order logic, that is, if there exists a SAT assignment consistent with these theories. 

Our encoding has several advantages over the one mentioned before.
First, we show that it is faster in practice.
Second, it benefits from the continuous efforts by fellow researchers on making SMT solvers more powerful.
Third, we argue that it is more natural, because it makes a distinction between the logic that is required to solve the problem, and the logic imposed by the background theories.
This allows us to easily extend the encoding to Moore and Mealy machines, and address a wider range of grammatical inference problems.

\section{Satisfiability Modulo Theories}
Let us first recall some basic terminology for \emph{propositional logic}.
In our presentation, we borrow notational conventions from \cite{NOT2006}.
Let $P$ be a set of \emph{Boolean variables}.
Such a variable $p \in P$ can be assigned either true or false.
A \emph{literal} $l$ is a variable $p$ or its negation $\neg p$.
A \emph{clause} $C$ is a disjunction of literals $l_1 \lor \ldots \lor l_n$.
A \emph{unit clause} is a clause consisting of a single literal.
The negation of a clause $C$ is a conjunction of the negations of its literals $\neg l_1 \land \ldots \land \neg l_n$.
A \emph{formula} $F$ is a conjunction of clauses $C_1 \land \ldots \land C_n$.
Finally, a (partial) \emph{assignment} $M$ is a (partial) mapping of variables to true or false.
A (partial) assignment can be seen as a conjunction of literals, and hence as a formula.

A clause $C$ is true in an assignment $M$ if at least one of its literals is in $M$. 
It is false in $M$ if all of the negations of its literals are in $M$.
Otherwise, it is undefined in $M$.
A formula $F$ is true in $M$, if all of its clauses are true in $M$.
In that case, $M$ is a \emph{model} of $F$.

\emph{Satisfiability} (SAT) is the problem of deciding for a given formula $F$ if there exists an assignment $M$ that is a model for $F$.
If no such assignment exists, then $F$ is called \emph{unsatisfiable}.

Before we give a description of satisfiability \emph{modulo theories}, let us briefly recall the necessary notions of \emph{first-order logic}.
We refer to \cite{KS2008} and \cite{NOT2006} for a more detailed explanation.
Central to first-order logic are the notions of \emph{formulae}, \emph{atoms}, \emph{terms} and \emph{variables}. 
First-order formulae are clauses constructed over atoms, which are in turn predicates constructed over Boolean variables and terms, which are in turn constructed over variables and constants that are defined over some (in our case finite) domain.
A \emph{theory} defines a set of valid formation rules for formulae, atoms and terms.
For our theories (\emph{linear inequality} and \emph{uninterpreted functions}), these rules are as follows.
\begin{itemize}
    \item A \emph{term} is inductively defined by the following rules:
        \begin{enumerate}
            \item Any \emph{variable} $x$ is a term.
            \item Any \emph{function} $f(t_1, \ldots, t_n)$ over terms $t_1, \ldots, t_n$ is a term.
        \end{enumerate}
    \item An \emph{atom} is a statement that may be true or false, depending on the value of its terms.
        It is defined as being either
        \begin{enumerate}
            \item an expression of the form $t = t'$ for terms $t$ and $t'$, or
            \item a \emph{predicate} $P(t_1, \ldots, t_n)$ over terms $t_1, \ldots, t_n$; specifically we are concerned with inequality relations ($<,>,\leq,\geq$).
        \end{enumerate}
    \item A \emph{formula} is inductively defined by the following rules:
        \begin{enumerate}
            \item A disjunction of formulae ($\lor$) is a formula.
            \item A negation ($\neg$) of a formula is a formula.
            \item An atom is a formula.
        \end{enumerate}
\end{itemize}

\emph{Equality logic for uninterpreted functions} is not concerned with the semantics for a function $f$.
This means that $f$ is not restricted by any axioms or rules of inference, if not explicitly added.
The theory does, however, impose the standard equality axioms over its terms (reflexivity, symmetry and transitivity). 
In addition, functions are required to be \emph{functionally consistent}, i.e.\
\[
    x_1 = y_1 \land \ldots \land x_n = y_n \implies f(x_1, \ldots, x_n) = f(y_1, \ldots, y_n)
\]

\emph{Satisfiability modulo theories} (SMT) is the problem of deciding satisfiability of a (conjunction of) first-order formula(e) with respect to one or more given theories.
Typically, this problem is solved as follows.
Initially, all atoms in the first-order formula(e) are replaced by Boolean variables, forgetting about the theory (therefore, we consider the same definitions and notation as given before for the propositional case).
The resulting (propositional) formula $F$ is presented to a SAT solver.
If a SAT solver determines it to be propositionally unsatisfiable, then we can conclude that the first-order formula is unsatisfiable as well.
If, instead, the SAT solver finds a propositional model $M$ for $F$, then $M$ is given to a so-called \emph{theory solver} that converts the model back to its first-order form, and checks if it is consistent with the theory.
If this is the case then we can conclude that $F$ is \emph{$T$-satisfiable} and that $M$ is a so-called \emph{$T$-model} of $F$.
Otherwise, the theory solver constructs a \emph{theory lemma}.
A theory lemma is a (propositional) clause $C$ that is a logical consequence from the theory. 
The SAT solver is then started again for the formula $F \land C$.
This process is repeated until the SAT solver finds a $T$-model for $F$, or concludes it is unsatisfiable.

Hence, SMT is the problem of deciding for a given theory $T$ and a given formula $F$, if $F$ is $T$-satisfiable, or equivalently, if there exists a $T$-model of $F$.
For a more detailed introduction to SMT we refer to \cite{MB2011}.
Improvements to the basic approach for solving SMT problems include: 
\begin{itemize}
    \item checking theory consistency of a (partial) assignment while it is being built,
    \item restarting the SAT solver from a point where the (partial) assignment was consistent with the theory,
    \item using a theory solver to guide the search for a satisfiable model by detecting literals that are a logical consequence of the theory, and
    \item periodically replacing a theory $T$ by a stronger theory $T \land T'$.
\end{itemize}
These recent improvements and ongoing advancements make SMT interesting from a grammatical inference perspective.

\section{Grammatical Inference as a SMT Problem}
\emph{Grammatical inference} is the study concerned with learning formal languages. 
One of the best studied problems in grammatical inference is that of finding a \emph{deterministic finite automaton} (DFA) of minimal size that accepts a given set of positive examples and rejects a given set of negative examples of an unknown regular language.

A DFA is a finite state machine that accepts and rejects \emph{strings}, which are sequences of \emph{symbols}.
Formally, we define it as a tuple $(\Sigma, Q, q_0, \delta, \lambda)$, where 
$\Sigma$ is a finite \emph{alphabet} of symbols,
$Q$ is a finite set of \emph{states},
$q_0$ is the \emph{initial state},
$\delta : Q \times \Sigma \to Q$ is a \emph{transition function} for states and symbols,
$\lambda : Q \to \mathbb{B}$ is an \emph{output function} that returns $\texttt{true}$ if a state is accepting, and $\texttt{false}$ if it is rejecting.

Let $x$ be a string, then we use $x_i$ to denote the symbol at the $i$th position of $x$.
The position $i$ of a symbol in a string satisfies $1 \leq i \leq |x|$ (i.e.\ the first symbol of $x$ is $x_1$).
We use $x_{[i,j]}$ to denote the substring of $x$ starting at position $i$ and ending at position $j$ (inclusive), i.e.\ $x = x_{[1,|x|]}$.

A DFA $D$ accepts a string $x$ if, starting in the initial state, the unique computation for $x$ ends in an accepting state.
This can be formalized as follows.
Let $x$ be a string over $\Sigma$, then $D$ accepts $x$ if a sequence of states $q'_0 \ldots q'_{|x|}$ exists such that
\begin{enumerate}
    \item $q'_0 = q_0$,
    \item $q'_{i} = \delta(q'_{i-1}, x_i)$ for $1 \leq i \leq |x|$, and
    \item $\lambda(q'_{|x|})=\texttt{true}$.
\end{enumerate}

Let $S_+$ be a \emph{sample} of strings that should be accepted, and let $S_-$ be a sample of strings that should be rejected.
Then one way of approaching the grammatical inference problem is by iteratively asking the following question.
\begin{quote}
    Is there a DFA of at most $n$ states that is \emph{consistent} with $S = \{S_+, S_-\}$, i.e.\ accepts all strings in $S_+$ and rejects all strings in $S_-$?
\end{quote}
If such a DFA $D$ of at most $n$ states exists, and one with at most $n-1$ states does not exist, then $D$ is a DFA of minimal size that is consistent with $S$.
The goal is to find such a consistent minimal DFA. 

\subsection{A natural encoding} \label{sec:natural}
This leads us to a natural encoding of the aforementioned problem in satisfiability modulo the theories of inequality and uninterpreted functions.
Let us consider the set of states of $D$ as a set of non-negative integers.
Then the following axioms assert that $D$ has at most $n$ states:
\begin{equation} \label{eq:natural-axiom-bool}
        \forall i \in \{0, \ldots, n-1\} \quad \forall a \in \Sigma \quad \bigvee\limits_{j=0}^{n-1} \delta(i, a) = j
\end{equation}
\begin{remark}
    The axioms in Equation~\ref{eq:natural-axiom-bool} can also be encoded in the following way (if the solver supports linear inequalities):    
    \begin{equation} \label{eq:natural-axiom-ineq}
        \forall i \in \{0, \ldots, n-1\} \quad \forall a \in \Sigma \quad \delta(i, a) \geq 0 \> \land \> \delta(i, a) < n
    \end{equation}
    
\end{remark}

If we assume without loss of generality that the initial state is $0$, then we can add the following constraints for the strings in $S_+$:
\begin{equation} \label{eq:natural-positive}
    \forall x \in S_+ \quad \lambda(\>\delta(\ldots \delta(\delta(0, x_1), x_2), \ldots x_{|x|})\>) = \texttt{true}
\end{equation}
Similarly, we can add the following constraints for the strings in $S_-$:
\begin{equation} \label{eq:natural-negative}
    \forall x \in S_- \quad \lambda(\>\delta(\ldots \delta(\delta(0, x_1), x_2), \ldots x_{|x|})\>) = \texttt{false} 
\end{equation}

The constraints and axioms in Equations~\ref{eq:natural-axiom-bool}-\ref{eq:natural-negative} are sufficient to determine if there is a DFA with at most $n$ states that is consistent with $S$.
If \emph{and only if} the resulting formula is unsatisfiable, such a DFA does not exist.
If it is satisfiable, however, the SMT solver's model provides us with $\delta$ and $\lambda$. 
Hence, we can construct a minimal consistent DFA $D$ for $S$ by iteratively incrementing $n$ (in Equation~\ref{eq:natural-axiom-bool} or \ref{eq:natural-axiom-ineq}).

Unfortunately, the nesting in the set of constraints given by Equations~\ref{eq:natural-positive} and \ref{eq:natural-negative} might make it hard for the theory solver to determine if the formula is consistent with the constraints given by Equation~\ref{eq:natural-axiom-bool} or \ref{eq:natural-axiom-ineq}.

\subsection{A more expressive encoding} \label{sec:expressive}
One solution to this is to define the constraints implied by strings in a non-nested way.
Similarly to Heule and Verwer \cite{HV2013}, and Bruynooghe et.\ al.\ \cite{BBB2015}, we use an \emph{augmented prefix tree} (APT) for this.
We, however, introduce a more concise set of constraints.
An APT $A$ can be considered a partial, tree-shaped DFA that is \emph{exactly consistent} with $S$, i.e.\ it accepts only the set $S_{+}$ and rejects only the set $S_{-}$.
For every state $q$ of $A$ there exists exactly one string that ends in $q$.
Therefore, we denote the unique state that a string $x$ ends in by $q_x$.
This implies that two strings $x$ and $y$ visit the same state if and only if they share a \emph{prefix}, i.e.\ $x_{[1,i]} = y_{[1,i]}$ for some $1 \leq i \leq \min(|x|, |y|)$.
We formally define an APT as $(\Sigma, Q, q_{\epsilon}, \delta, Q_S, \lambda)$, where 
$\Sigma$ and $Q$ are the same as before, 
$q_{\epsilon}$ is the initial state (i.e.\ the state reached by the empty string $\epsilon$),
$\delta: Q \times \Sigma \rightharpoonup Q$ is a \emph{partial transition function} that satisfies the aforementioned property, 
$Q_S = \{q_x \in Q : x \in S\}$ is the subset of $Q$ in which a string of $S$ ends, and 
$\lambda: Q_S \to \mathbb{B}$ is an output function for these states.
Indeed, no output is defined for the states in $Q \setminus Q_S$.


Let us define the set of constraints for finding a DFA $D = (\Sigma, Q^D, q_0^D, \delta^D, \lambda^D)$ that is consistent with an APT $A = (\Sigma, Q^A, q_{\epsilon}^A, \delta^A, Q_S^A, \lambda^A)$ for a given sample $S = \{S_+, S_-\}$. 
Recall that such a DFA is consistent if and only if it accepts all strings in $S_+$ and rejects all strings in $S_-$,  i.e.\ for each $x$ in $S$ $\lambda^D(\delta^D(q_0, x)) = \lambda^A(q_x)$ (we slightly abuse notation here by extending $\delta$ to strings). 
Such a DFA has at most as many states as the APT (but typically significantly less).
Therefore, there must exist a surjective (i.e.\ many-to-one) function $\pi: Q^A \to Q^D$.
Our goal is to find a set of constraints for $\pi$ that make sure that our target DFA $D$ is consistent.

First, let us encode the (partial) transition function $\delta^A$: 
\begin{equation} \label{eq:expressive-transitions}
    \forall q_{xa} \in Q^A : x \in \Sigma^{\ast}, a \in \Sigma \quad \delta^D(\pi(q_x), a) = \pi(q_{xa})
\end{equation}
Now, let us encode the output function $\lambda^A$:
\begin{equation} \label{eq:expressive-output}
    \forall q \in Q_S^A \quad \lambda^D(\pi(q_x)) = \lambda^A(q_x) 
\end{equation}

The problem at hand is to find a `smallest' $\pi$ function; i.e.\ there should be no other function with a smaller image that satisfies these constraints.
We can encode this as follows.
Let us (again) consider the set of states of $D$ as a set of non-negative integers.
Then the following axioms assert that $D$ has at most $n$ states:
\begin{equation} \label{eq:expressive-axiom-bool}
        \forall q \in Q^A \quad \bigvee\limits_{i=0}^{n-1} \pi(q) = i
\end{equation}
\begin{remark}
    The axioms in Equation~\ref{eq:expressive-axiom-bool} can also be encoded in the following way (if the solver supports linear inequalities):
    \begin{equation} \label{eq:expressive-axiom-ineq}
        \forall q \in Q^A \quad \pi(q) \geq 0 \> \land \> \pi(q) < n
    \end{equation}
\end{remark}

Again, these constraints and axioms are sufficient to determine if there is a DFA with at most $n$ states that is consistent with $S$, and if the resulting formula is satisfiable the SMT solver's model provides us with $\delta^D$ and $\lambda^D$. 
Hence, we can construct a minimal consistent DFA $D$ for $S$ by iteratively incrementing $n$ in Equation~\ref{eq:expressive-axiom-bool} or \ref{eq:expressive-axiom-ineq}.

\subsection{An extension for Moore and Mealy machines} \label{sec:transducers}
An advantage of the encoding presented in Section~\ref{sec:expressive} (as opposed to the one presented in Section~\ref{sec:natural}) is that it can easily be extended to learn Moore and Mealy machines.
In this section we present such an encoding.

A \emph{Moore machine} is a \emph{transducer} that produces an \emph{output symbol} in each state. 
Formally, it is a tuple $(\Sigma, \Lambda, Q, q_0, \delta, \lambda)$ where
$\Sigma$, $Q$, $q_0$ and $\delta$ are the same as for a DFA,
$\Lambda$ is a finite alphabet of output symbols, and
$\lambda : Q \to \Lambda$ is an output function.
A Moore machine produces an output symbol every time it (re-)enters a state.
Therefore, a sample $S$ for a Moore machine consists of \emph{traces}, which are pairs $(x, y)$ where $x = x_1 \ldots x_{|x|}$ is a string over $\Sigma$ and $y = y_0 y_1 \ldots y_{|x|}$ is a string over $\Lambda$ (observe that $|y| = |x| + 1$).

A \emph{Mealy machine} is a transducer whose output symbols are determined by both its current state and the current input symbol.
Formally, it is the same as a Moore machine, except that $\lambda : Q \times \Sigma \to \Lambda$ is a \emph{transition output function}.
A Mealy machine produces an output every time it makes a transition.
A sample $S$ for a Mealy machine consists of traces $(x, y)$ where $x = x_1 \ldots x_{|x|}$ is a string over $\Sigma$ and $y = y_1 \ldots y_{|x|}$ is a string over $\Lambda$ (observe that $|y| = |x|$).

It has been shown in \cite{HU1979} that Moore and Mealy machines are equi-expressive if we neglect the output produced by the initial state of a Moore machine.  
Therefore, we can define an APT for a sample $S$ of traces produced by either a Moore or Mealy machine $M = (\Sigma, \Lambda, Q^M, q_0^M, \delta^M, \lambda^M)$ in a similar way.
We choose to define it as a tuple $A = (\Sigma, \Lambda, Q^A, q_{\epsilon}^A, \delta^A, \lambda^A)$, where $\Sigma$, $\Lambda$, $Q^A$, $q_{\epsilon}^A$ and $\delta^A$ are as you would expect from previous definitions, and $\lambda^A : Q^A \setminus q_{\epsilon}^A \to \Lambda$ is an output function that is defined for all states except $q_{\epsilon}^A$.
One can afterwards add $\lambda^A(q_{\epsilon}^A) = y_0$ for any output string $y$ in $S$ if $M$ is a Moore machine.

We can now determine if there is a Moore or Mealy machine with at most $n$ states that is consistent with $S$ by using the set of constraints and axioms presented in Section~\ref{sec:expressive}, if we
\begin{enumerate}
    \item define $\pi : Q^A \to Q^M$ accordingly,
    \item replace $D$ with $M$ in Equation~\ref{eq:expressive-transitions}, and
    \item replace Equation~\ref{eq:expressive-output} with Equation~\ref{eq:moore-output} (for Moore machines) or \ref{eq:mealy-output} (for Mealy machines).
\end{enumerate}

\begin{equation} \label{eq:moore-output}
    \forall q \in Q^A \quad \lambda^M(\pi(q)) = \lambda^A(q) 
\end{equation}
\begin{equation} \label{eq:mealy-output}
    \forall q_{xa} \in Q^A : x \in \Sigma^{\ast}, a \in \Sigma \quad \lambda^M(\pi(q_x), a) = \lambda^A(q_{xa}) 
\end{equation}

\section{Experimental Results}
We have implemented our encodings using the Python front-end of Z3 \cite{MB2008}, and we have conducted some initial experiments that assess the scalability of the different encodings and their applicability in practice\footnote{See \url{https://gitlab.science.ru.nl/rick/z3gi}.}.

In our first set of experiments we benchmark the running times of the following encodings:
\begin{itemize}
    \item Heule and Verwer's propositional encoding without redundant constraints \cite[Table~1]{HV2013} (HV-1),
    \item Heule and Verwer's propositional encoding with redundant constraints \cite[Table~1]{HV2013} (HV-2),
    \item Section~\ref{sec:natural} with Equation~\ref{eq:natural-axiom-bool} (\ref{sec:natural}-\ref{eq:natural-axiom-bool}),
    \item Section~\ref{sec:natural} with Equation~\ref{eq:natural-axiom-ineq} (\ref{sec:natural}-\ref{eq:natural-axiom-ineq}),
    \item Section~\ref{sec:expressive} with Equation~\ref{eq:expressive-axiom-bool} (\ref{sec:expressive}-\ref{eq:expressive-axiom-bool}), and
    \item Section~\ref{sec:expressive} with Equation~\ref{eq:expressive-axiom-ineq} (\ref{sec:expressive}-\ref{eq:expressive-axiom-ineq}).
\end{itemize}

For this purpose, we define a DFA $D_k^{\textrm{mod}} = (\Sigma, Q, q_0, \delta, \lambda)$ that accepts strings $x$ if $|x| \bmod k = 0$:
\begin{itemize}
    \item $\Sigma = \{\mathtt{a}\}$,
    \item $Q = \{ i \in \mathbb{N} : 0 \leq i < k \}$,
    \item $q_0 = 0$,
    \item $\delta(i, \mathtt{a}) = (i+1) \bmod k$ for $i$ in $Q$, and
    \item $\lambda(0) = \texttt{true}$ and $\lambda(i) = \texttt{false}$ for $1 \leq i < k$.
\end{itemize}
This DFA $D_k^{\textrm{mod}}$ is then used to construct a benchmark sample $S_k^{\textrm{mod}}$ for $1 \leq k \leq 12$ that contains all strings up to length 100 (i.e.\ $\Sigma^{\leq 100}$).

\begin{figure}
    \centering
    \begin{tikzpicture}
        \begin{semilogyaxis}[log basis x={10}, xlabel={$k$}, every axis x label/.style={at={(ticklabel* cs:1)},anchor=north west}]
            \addplot[color=gray, solid] table[x=states, y=pr_nr] {mod100.table};
            \addplot[color=gray, dashed] table[x=states, y=pr_re] {mod100.table};
            \addplot[thick, solid] table[x=states, y=na_qu_ni] {mod100.table};
            \addplot[thick, dashed] table[x=states, y=na_qu_in] {mod100.table};
            \addplot[thick, dotted] table[x=states, y=ex_qu_ni] {mod100.table};
            \addplot[thick, dashdotted] table[x=states, y=ex_qu_in] {mod100.table};
        \end{semilogyaxis}
    \end{tikzpicture}
    \label{fig:mod-experiments}
    \caption{Running times (in seconds) for HV-1 (gray solid), HV-2 (gray dashed), \ref{sec:natural}-\ref{eq:natural-axiom-bool} (solid), \ref{sec:natural}-\ref{eq:natural-axiom-ineq} (dashed), \ref{sec:expressive}-\ref{eq:expressive-axiom-bool} (dotted) and \ref{sec:expressive}-\ref{eq:expressive-axiom-ineq} (dashdotted)}
\end{figure}
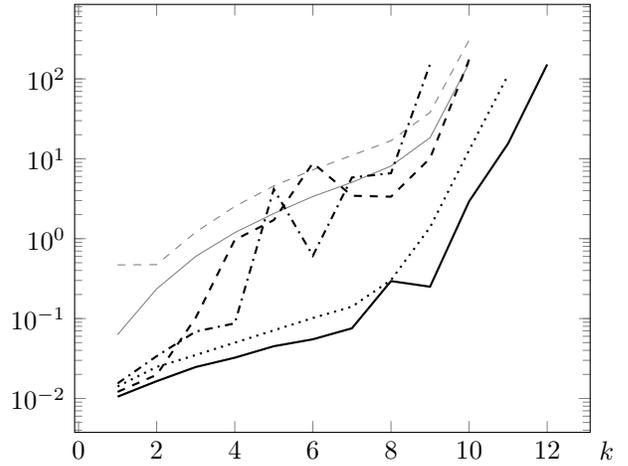

Results for the experiments are shown in Figure~\ref{fig:mod-experiments}.
The experiments were performed on an Intel Core i5-4258U processor and the timeout for an experiment was set to 10 minutes.
Each experiment was repeated 5 times.
Interestingly, the `outliers' occurred consistently in these experiments.

In our second set of experiments we simulate the case study from \cite{ARP2013}.
Here, the authors use model learning \cite{Vaa2017} techniques to obtain Mealy machines for different bank cards\footnote{See \url{http://automata.cs.ru.nl/BenchmarkBankcard}.}.
We first use the Wp-method \cite{FBK1991} to generate a characterizing sample for these Mealy machines.
Then, we use the following encodings to reconstruct the Mealy machine:
\begin{itemize}
    \item Section~\ref{sec:transducers} with Equations~\ref{eq:expressive-axiom-bool} and \ref{eq:mealy-output} (\ref{sec:transducers}-\ref{eq:expressive-axiom-bool}), and
    \item Section~\ref{sec:transducers} with Equations~\ref{eq:expressive-axiom-ineq} and \ref{eq:mealy-output} (\ref{sec:transducers}-\ref{eq:expressive-axiom-ineq}).
\end{itemize}
The results of the experiments are shown in Table~\ref{tab:bankcard-experiments}.
Here, $||S|| = \sum_{x \in S} |x|$.
As expected, the learned models were equivalent to the source Mealy machines.

\begin{table}
    \centering
    \caption{Running times and details for bank card experiments}
    \begin{tabular}{l l l l l l l l}
        \toprule
        Bank card & $|\Sigma|$ & $|\Lambda|$ & $|Q|$ & $|S|$ & $||S||$ & \ref{sec:transducers}-\ref{eq:expressive-axiom-bool} & \ref{sec:transducers}-\ref{eq:expressive-axiom-ineq} \\
        \midrule
        Maestro & 14 & 10 & 6 & 188 & 740 & 0.651s & 3.525s \\
        MasterCard & 14 & 9 & 6 & 226 & 940 & 0.661s & 3.111s \\
        PIN & 14 & 10 & 6 & 255 & 1022 & 0.738s & 7.077s \\
        SecureCode & 14 & 9 & 4 & 103 & 371 & 0.094s & 0.498s \\
        VISA & 15 & 11 & 9 & 403 & 1835 & 9.523s & 289.378s \\
        \bottomrule
    \end{tabular}
    \label{tab:bankcard-experiments}
\end{table}

\section{Conclusions and Future Work}
We have presented an encoding in SMT for learning a minimal consistent DFA, Moore machine or Mealy machine from a set of sequences of symbols.
Our experimental results show that these encodings improve upon the state-of-the-art, and are useful in practice. 
The time required to learn a model, however, grows rapidly in its number of states.
In the future we wish to address this problem and extend this work by:
\begin{enumerate}
    \item defining a counterexample-driven algorithm that incrementally refines a minimal consistent model,
    \item combining our approach with existing learning algorithms, and
    \item extending the encoding for richer model formalisms, such as register automata.
\end{enumerate}

\section*{Acknowledgments}
This work is supported by the Netherlands Organisation for Scientific Research (NWO) project 628.001.009 on Learning Extended State Machine for Malware Analysis (LEMMA).



\bibliographystyle{IEEEtran}
\bibliography{IEEEabrv,lib}
%
%
%

\end{document}